\documentclass
[twocolumn,aps,prc,amsmath,amssymb,floatfix]
{revtex4-1}
\usepackage{CJK}
\usepackage[dvips]{graphicx}
\usepackage{bm}                      
\usepackage{mathptmx}                
\usepackage{dcolumn}                 
\usepackage{multirow}                
\usepackage{xcolor}                  %
\graphicspath{{fig/}}

\begin{document}

\def\bea{\begin{eqnarray}} \def\eea{\end{eqnarray}}
\def\be{\begin{equation}} \def\ee{\end{equation}}
\def\bal#1\eal{\begin{align}#1\end{align}}
\def\bse#1\ese{\begin{subequations}#1\end{subequations}}
\def\rra{\right\rangle} \def\lla{\left\langle}
\def\rv{\bm{r}} \def\tv{\bm{\tau}} \def\sv{\bm{\sigma}}
\def\al{\alpha}
\def\eps{\varepsilon}
\def\ms{M_\odot}
\def\esym{E_\text{sym}}
\def\mmax{M_\text{max}}
\def\prat{p_\text{ratio}}
\def\mrat{M_\text{ratio}}
\def\tt{(\tv_1\cdot\tv_2)} \def\ss{(\sv_1\cdot\sv_2)}
\def\pt{p_\text{th}} 
\def\mev{\;\text{MeV}}
\def\fm3{\;\text{fm}^{-3}}
\long\def\hj#1{\color{blue}#1\color{black}}

\title{
Microscopic nuclear equation of state at finite temperature
and stellar stability
}

\begin{CJK*}{UTF8}{gbsn}

\author{Hong-Ming Liu (刘宏铭)$^1$} 
\author{Jing Zhang (张晶)$^1$} 
\author{\hbox{Zeng-Hua Li (李增花)$^1$}} \email[]{zhli09@fudan.edu.cn}
\author{Jin-Biao Wei (魏金标)$^2$} 
\author{G. F. Burgio$^3$}
\author{H.-J. Schulze$^3$}

\affiliation{$^1$%
\hbox{Institute of Modern Physics,
Key Laboratory of Nuclear Physics and Ion-beam Application (MOE),
Fudan University, Shanghai 200433, P.R.~China}\\
\hbox{$^2$Physics Department, University of Geosciences, Wuhan, P.R.~China}\\
\hbox{$^3$INFN Sezione di Catania, Dipartimento di Fisica,
Universit\'a di Catania, Via Santa Sofia 64, 95123 Catania, Italy}
}

\date{\today}

\begin{abstract}
A microscopic nuclear equation of state
compatible with all current astrophysical constraints
constructed within the Brueckner-Hartree-Fock formalism is presented
and extended in a consistent way to finite temperature.
The effects of finite temperature
on the properties of neutron stars
are studied in detail
and a universal relation regarding stellar stability is proposed.
\end{abstract}

\maketitle
\end{CJK*}

\section{Introduction}

The determination of the nuclear equation of state (EOS) for neutron stars (NSs)
currently represents a formidable theoretical challenge in nuclear astrophysics.
In fact NSs are characterized by a density which spans over fourteen orders
of magnitude from the crust to the inner core,
where significant isospin asymmetries can be present.
However, whereas for cold NSs in weak beta-equilibrium
the EOS reduces to a relation between pressure and energy density,
the dynamical evolutions of core-collapse supernovae (CCSN)
\cite{Burrows13,Constantinou14,Janka16,Mueller20},
proto-NSs
\cite{Burrows86,Prakash97,Pons99,Nicotra06,Li10,Burgio11},
and binary neutron star (BNS) mergers
\cite{Paschalidis12,Kaplan14,Marques17,Lalit19,Beznogov20,
Figura20,Figura21,Raithel21,Koliogiannis21},
require an EOS strongly dependent on the temperature $T$,
which can rise in some cases up to a few tens or even above a hundred MeV.
Nuclear physics experiments in terrestrial laboratories cannot explore
the high-density regions encountered in NSs,
and therefore theoretical methods have to be devised for the exploration
of this regime.
While the subject remained rather academic for several years,
the recent progress of astrophysical observations
opened the possibility that theoretical speculations could be confronted
with observational data.
For this purpose many theoretical methods have been devised,
see, e.g., Refs.~\cite{Lattimer16,Oertel17,Burgio18a,Burgio21,Stone21}
for recent reviews.

From the observational point of view,
NS mass measurements are particularly relevant because they provide
lower limits on the maximum gravitational mass of stable configurations.
The observations of pulsars with masses around $2\ms$,
which include PSR J1614-2230 ($M=1.97\pm0.04\;\ms$) \cite{Demorest10},
and PSR J0348+0432 ($M=2.01\pm0.04\;\ms$) \cite{Antoniadis13},
placed some of the most stringent constraints on the high-density EOS so far.

Gravitational waves emitted during the late inspiral phase
of the BNS merger event GW170817 have allowed to determine
a combined tidal deformability of the two NSs
\cite{Abbott17,Abbott18,Abbott20,Baiotti19}.
Along with the detection of its electromagnetic counterparts,
AT2017gfo \cite{knova1,knova2,knova3,knova4,knova5},
this allowed to constrain the EOS of dense matter by ruling out
very stiff EOSs due to an upper limit on the average tidal deformability.
A lower limit on the tidal deformability could also be extracted
by considering the large amount of ejected matter,
which powers the kilonova AT2017gfo \cite{Radice2018}.

The Neutron Star Interior Composition Explorer (NICER) mission
has recently provided two simultaneous mass and radius measurements
for PSR J0030+0451 with
$R(1.44^{+0.15}_{-0.14}\ms)=13.02^{+1.24}_{-1.06}\;$km \cite{Miller19} and
$R(1.34^{+0.15}_{-0.16}\ms)=12.71^{+1.14}_{-1.19}\;$km \cite{Riley19}
and for J0740+6620 with
$R(2.08 \pm 0.07\ms)=13.7^{+2.6}_{-1.5}\;$km \cite{Miller21} and
$R(2.072^{+0.067}_{-0.066}\ms)=12.39^{+1.30}_{-0.98}\;$km \cite{Riley21}.
The difference between these estimates reflects the model dependence
of the experimental analyses.
Combining the NICER results with the limits on the NS maximum mass
as well as the tidal deformability from GW170817 \cite{Abbott20}
leads to constraints on the beta-equilibrated EOS
for densities in the range
$1.5 \rho_0 \lesssim \rho_B \lesssim 3\rho_0$
(being $\rho_0$ the nuclear saturation density)
\cite{Miller21,Raaijmakers21},
which can be translated into limits for NS radii
\cite{Miller21,Riley19,Miller19,Riley21,Miller21,Pang21,Raaijmakers21}.

Theoretical EOSs are now routinely confronted with these new data,
which has already lead to the exclusion of many EOSs that do
not fulfill the current constraints on NS mass and radius,
see, e.g., \cite{Wei21}.
The theoretical modeling of the EOS can therefore benefit from these data,
and this allows to extend the preparation of theoretical nuclear EOSs
in a controlled way to more exotic conditions,
in particular finite temperature.
So far only a few theoretical EOSs are available for this purpose \cite{Wei21},
because useful observational constraints regarding temperature effects
are still not available.
In particular, among the ab-initio theoretical calculations we mention the
Brueckner-Hartree-Fock (BHF) EOSs V18 and N93
\cite{Li08a,Li08b,Burgio10,Lu19,Wei21},
based on the Bloch-De Dominicis approach,
and those based on the variational APR EOS \cite{Akmal98},
i.e., TNTYST \cite{Togashi17} and SRO(APR) \cite{Schneider19}.
Other approaches based on the covariant density-functional theory are available,
and among those we mention
Shen11 \cite{Shen98,Shen11},
Shen20 \cite{Shen20},
HS(DD2) \cite{Hempel10,Typel10},
SFHx \cite{Hempel10,Steiner13},
and FSU2H \cite{Tolos17,Tolos17b}.
Also non-relativistic Skyrme-type EOSs at finite temperature are widely used,
in particular the Lattimer-Swesty EOS \cite{Lattimer91}.
However, the current status of theoretical investigation is expected to change
in the near future,
in particular once features of the hot remnant of BNS merger events
will be revealed by more sensitive next-generation gravitational-wave detectors
\cite{Maggiore20}.

This work is dedicated to the construction of a theoretical nuclear EOS
compatible with all current constraints on the cold EOS at high density,
and extended in a microscopic way to relevant astrophysical temperatures.
We continue several previous works regarding the EOS based on the BHF approach
with an approximate treatment of finite temperature, i.e.,
the Frozen Correlations Approximation (FCA)
for the single-particle (s.p.) potential,
in which the correlations at finite temperature are assumed to be the same
as at $T=0$
\cite{Zhou04,Li08a,Li08b,Burgio18,Lu19,Wei20,Figura20,Figura21,Li21,Wei21}.
This allowed to solve the equations for the in-medium scattering $K$ matrix
only at $T=0$.
In this work we proceed to discuss results obtained by
introducing temperature effects into the $K$-matrix equation,
and compare with the FCA.

Our paper is organized as follows.
In Sec.~\ref{s:bhf} we briefly discuss the theoretical formalism
of the finite-temperature BHF approach,
and the EOS for symmetric and pure neutron matter at finite temperature.
Useful fits are given, which may help in the calculation of the free energy,
composition, and EOS of stellar matter.
In Sec.~\ref{s:res} numerical results regarding thermal effects
on asymmetric and beta-stable matter are presented,
including the proton fraction, the maximum mass,
and possible correlations with thermodynamical variables.
Conclusions are drawn in Sec.~\ref{s:end}.

\section{Formalism}
\label{s:bhf}

We review briefly the construction of the finite-temperature EOS
and its application to stellar structure.

\subsection{Brueckner-Bethe-Goldstone theory at finite temperature}
\label{s:bhft}

All thermodynamic quantities of hot stellar matter of our interest
can be derived from the free energy density,
which consists of two contributions,
\be
 f = f_N + f_L \:,
\label{e:f}
\ee
where $f_N$ is the nucleonic part
and $f_L$ denotes the contribution of noninteracting leptons
$e,\mu,\nu_e,\nu_\mu$,
and their antiparticles.
In this work,
we employ the BHF approach for asymmetric nuclear matter at finite temperature
\cite{Bloch58,Bloch59a,Bloch59b,Lejeune86,Bombaci96,Baldo99a,Baldo99b,
Zuo04,Burgio10,Logoteta21}
to calculate the nucleonic contribution.
The essential ingredient of this approach is the
in-medium interaction matrix $K$,
which satisfies the self-consistent equations
\be
  K(\rho,x_p;E) = V + V \;\text{Re} \sum_{1,2}
 \frac{|12 \rangle (1-n_1)(1-n_2) \langle 1 2|}
 {E - e_1-e_2 +i0} K(\rho,x_p;E) \:
\label{e:BG}
\ee
and
\bea
 e_1 &=& \frac{k_1^2}{2m_1} + U_1 \:,
\\
 U_1(\rho,x_p) &=& \sum_2 n_2
 \langle 1 2| K(\rho,x_p;e_1+e_2) | 1 2 \rangle_a
\label{e:uk}
\eea
for the s.p.~energy $e_1$ and potential $U_1$.
Here $E$ is the starting energy and
$x_p=\rho_p/\rho$ is the proton fraction, with
proton and total baryon densities $\rho_p$ and $\rho$, respectively.
The multi-indices 1,2 denote in general momentum, isospin, and spin.

In this work the input nucleon-nucleon ($NN$) interaction $V$ is
the Argonne $V_{18}$ potential \cite{Wiringa95} supplemented by
compatible three-body forces (TBF),
obtained employing the same meson-exchange parameters as the two-body potential
\cite{Grange89,Zuo02,Li08a}.
They have a strongly repulsive character at high density,
at variance with the phenomenological Urbana TBF (UIX)
adopted earlier \cite{Burgio10},
and the closely related chiral TBF of order N2LO
\cite{Li12,Piarulli16,Bombaci18,Logoteta21}.
The TBF are reduced to an effective, density-dependent,
two-body force by averaging over the third nucleon in the medium,
\be
 \overline{V}_{\!\!12}(\rv_{12}) =
 \rho \int\! d^3\!\rv_3 \sum_{\sigma_3,\tau_3}
 g_{13}(r_{13})^2 g_{23}(r_{23})^2\, V_{132} \:,
\label{e:va}
\ee
the average being weighted by the BHF correlation function
$g_\al(r) = 1-\eta_\al(r)$,
related to the $r$-space defect functions in the partial waves $\al$,
\bea
 \eta_\al(r) &=& j_\al(r) - u_\al(r)
 \:,
\label{e:eta}
\eea
with the free wave function (Bessel function) $j_\alpha$,
and the correlated wave function $u_\alpha$,
which takes into account the nucleon-nucleon in-medium correlations
\cite{Blatt75,Coon79,Grange89,Baldo99a,Zuo02,Zhou04,Li08a,Li16,Lu18}.
This produces an effective two-nucleon potential with operator structure
\bea
 \overline{V}_{\!\!12}(\rv) &=&
 \tt\ss V_C(r) + \ss V_S(r) + V_I(r)
\nonumber\\&&
  +\, S_{12}(\hat{\rv}) \big[ \tt V_T(r) + V_Q(r) \big] \:,
\label{e:vavm}
\eea
where $S_{12}(\hat{\rv}) =
3(\sv_1 \cdot \hat{\rv})(\sv_2 \cdot \hat{\rv}) - \sv_1 \cdot \sv_2$
is the tensor operator
and the five components
$V_O,\; O=C,S,I,T,Q$
depend on the nucleon densities $\rho_n$ and $\rho_p$.


At finite temperature,
$n(k)$ in Eqs.~(\ref{e:BG}) and (\ref{e:uk}) is a Fermi distribution.
For a given density and temperature, these equations have
to be solved self-consistently along with the equations for
the auxiliary chemical potentials $\tilde{\mu}_{n,p}$,
\be
 \rho_i = 2\sum_k n_i(k) =
 2\sum_k {\left[\exp{\Big(\frac{e_i(k)-\tilde{\mu}_i}{T}\Big)}
 + 1 \right]}^{-1} \:.
\label{e:rho}
\ee
Within this approximation,
the nucleonic free energy density is \cite{Lejeune86,Baldo99a,Baldo99b}
\be
 f_N = \sum_{i=n,p} \left[ 2\sum_k n_i(k)
 \left( \frac{k^2}{2m_i} + \frac{1}{2}U_i(k) \right) - Ts_i \right] \:,
\label{e:fn}
\ee
where
\be
 s_i = - 2\sum_k \Big( n_i(k) \ln n_i(k) + [1-n_i(k)] \ln [1-n_i(k)] \Big)
\label{e:entr}
\ee
is the entropy density for the component $i$ treated as a free Fermi gas with
spectrum $e_i(k)$.

All necessary thermodynamical quantities are derived
in a consistent way from the total free energy density $f$,
namely the chemical potentials $\mu_i$,
pressure $p$,
internal energy density $\eps$, and
entropy density $s$
are
\bea
 \mu_i &=& \frac{\partial f}{\partial \rho_i}
 \Big|_{T,\{\rho_j\}_{j \neq i}} \:,
\label{e:mui}
\\
 p &=& \rho^2 \frac{\partial{(f/\rho)}} {\partial{\rho}}\Big|_T
 = \sum_i \mu_i \rho_i - f \:,
\label{e:eosp}
\\
 \eps &=& f + Ts \:,\quad
 s = -\frac{\partial f}{\partial T}\Big|_{\{\rho_i\}} \:.
\label{e:eoss}
\eea

\begin{figure*}[t]
\vskip-4mm
\centerline{\hskip-12mm\includegraphics[scale=0.38,angle=0,clip]{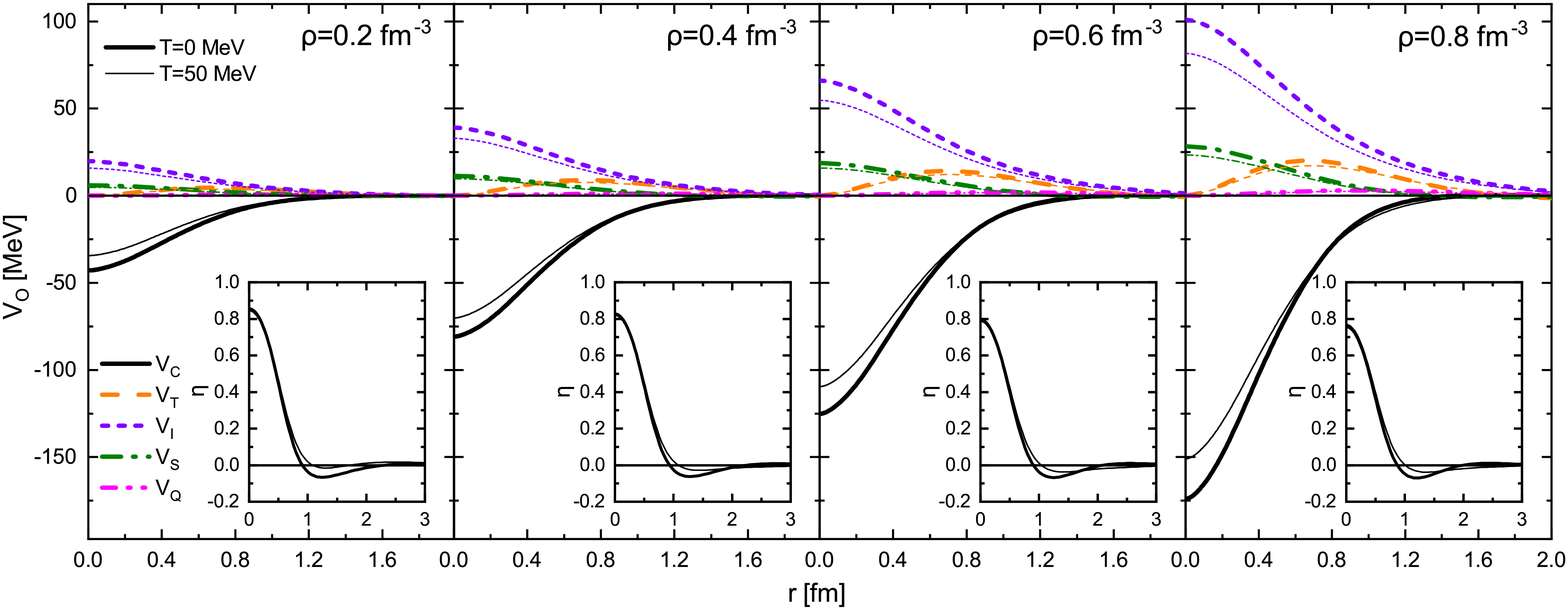}}
\vskip-6mm
\caption{
The averaged TBF components, Eq.~(\ref{e:vavm}),
as a function of the radial distance
in symmetric nuclear matter
for different values of the nucleon density $\rho$
at $T=0$ (thick curves) and $T=50\mev$ (thin curves).
${^1S_0}$ defect functions are shown in the insets.
See text for details.}
\label{f:tbf}
\end{figure*}

In order to simplify and speed up the numerical procedure,
we previously often employed the FCA
\cite{Baldo99b,Nicotra06,Burgio11b},
in which the s.p.~potentials $U_i(k)$ at finite temperature
are approximated by the ones calculated at $T=0$,
and thus the Bethe-Goldstone equation~(\ref{e:BG}) needs only be solved at $T=0$.
This means that the correlations at $T \neq 0$
are assumed to be essentially the same as at $T=0$.
In this paper we go beyond the FCA,
and solve the system of equations (\ref{e:BG}) to (\ref{e:rho})
at finite temperature.

It should be noted that while the two-body potential $V$
is considered temperature independent,
there are some notable effects of temperature on the averaged TBF, namely
the defect functions, Eq.~(\ref{e:eta}),
change slightly at finite temperature due to the
$T$ dependence of the $K$~matrix.
This is illustrated in Fig.~\ref{f:tbf},
where the five TBF components, Eq.~(\ref{e:vavm}),
and the defect function in the $\alpha={^1S_0}$ channel (insets)
are plotted as functions of the radial distance,
for SNM at several values of the nucleon density, 
comparing results at $T=0$ and $T=50\mev$.
(The results in beta-stable matter are very similar). One can clearly see that the five components
are all weakened at finite temperature,
which is due to an increase of the defect functions
and thus a reduction of the correlation functions~$g$ in Eq.~(\ref{e:va}).
In other words,
at finite $T$ interacting nucleons are kept at larger distance
and the effective `core' of the $NN$ interaction increases.
In conclusion, the TBF turn out to be weakened in hot matter.
This has consequences for the EOS, as discussed in the next section.

\subsection{EOS for symmetric and pure neutron matter}
\label{s:eosT}

\begin{figure}[t]
\vspace{-10mm}
\centerline{\includegraphics[scale=0.48,angle=0,clip]{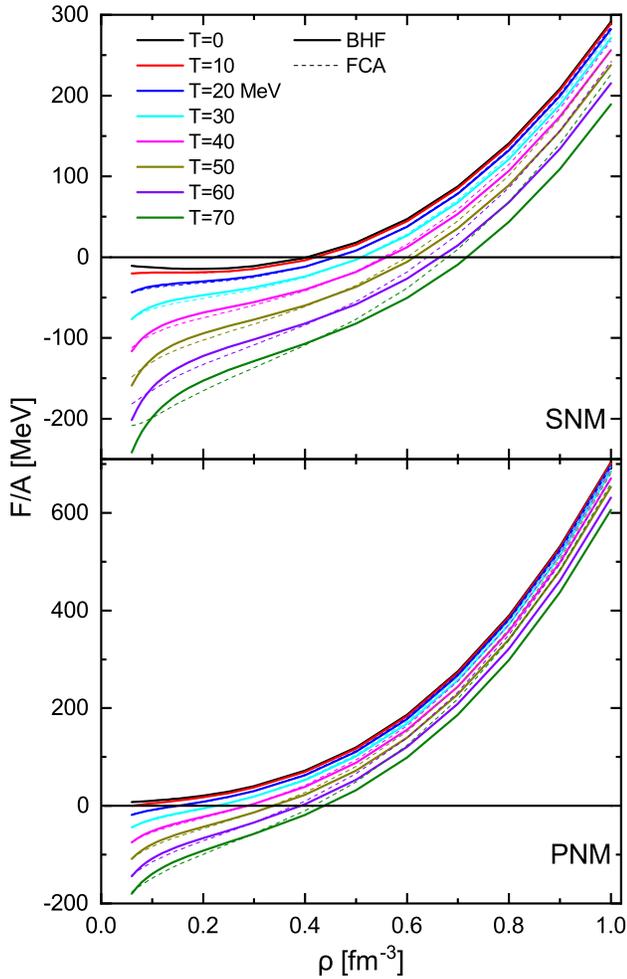}}
\vspace{-15mm}
\caption{
Free energy per nucleon as a function of nucleon density
for symmetric (upper panel) and pure neutron matter (lower panel).
The temperatures vary from 0 to $70\mev$ in steps of $10\mev$.
Results for BHF (solid curves) and FCA (dashed curves) procedures are compared.
}
\label{f:fa}
\end{figure}

\begin{table}[t]
\caption{
Parameters of the fit for the free energy per nucleon $F/A$,
Eq.~(\ref{e:fitf}),
for symmetric nuclear matter (SNM) and pure neutron matter (PNM)
in BHF and FCA procedures.
}
\def\myc#1{\multicolumn{1}{c}{$#1$}}
\def\myt#1{\multicolumn{1}{c}{$\tilde{#1}$}}
\renewcommand{\arraystretch}{1.2}
\begin{ruledtabular}
\begin{tabular}{lr|rrrr|rrrrr}
     &     & \myc{a} & \myc{b} & \myc{c} & \multicolumn{1}{c|}{$d$}
           & \myt{a} & \myt{b} & \myt{c} & \myt{d} & \myt{e} \\
\hline
\multirow{2}{*}
 {BHF}& SNM & -57 & 356 & 2.71 & -8 & -203 & 135 & -43 &  48 & 2.63 \\
      & PNM &  45 & 654 & 2.83 &  5 & -195 & 104 & -67 &  68 & 2.25 \\
\hline
\multirow{2}{*}
 {FCA}& SNM & -57 & 356 & 2.71 & -8 & -144 & 210 & -37 &  60 & 2.59 \\
      & PNM &  45 & 654 & 2.83 &  5 &  -92 & 150 & -34 &  43 & 2.45 \\
\end{tabular}
\end{ruledtabular}
\label{t:fit}
\end{table}

According to the formalism just illustrated,
we start by displaying in Fig.~\ref{f:fa} the free energy per nucleon,
$F/A=f_N/\rho$,
as a function of the nucleon density $\rho$,
for symmetric nuclear matter ($x_p=1/2$, SNM)
and pure neutron matter ($x_p=0$, PNM),
for several values of temperature between 0 and $70\mev$.
Results of the FCA (dashed curves)
and the exact calculations
(solid curves, labeled BHF here and throughout this paper)
are shown.
At $T=0$ the free energy coincides with the internal energy
and the corresponding SNM curve is the usual nuclear matter saturation curve.
For PNM the temperature effect is less pronounced due to the larger
Fermi energy and free energy of the neutrons at a given density.
Comparing BHF and FCA procedures,
one notes differences at low density and high temperature
(caused by a great importance of temperature effects $\sim T/e_F$)
as well as at high density,
where the reduction of the TBF strength causes a clearly visible
softening of $F/A$
at $\rho\gtrsim 0.4\fm3$ for both SNM and PNM in the BHF procedure.

This effect will not be seen when disregarding temperature effects on the TBF,
as in the FCA
or for example in the recent results \cite{Logoteta21}
obtained within the same BHF approach at finite temperature
using a chiral two-nucleon potential with consistent N2LO$\Delta$ TBF
\cite{Piarulli16,Bombaci18},
which produces a much softer EOS with respect to our V18+TBF model.

As in our previous publications we provide
analytical fits of the free energy $F/A(\rho,T)$
for SNM/PNM and BHF/FCA
according to the following functional form
\bea
 \frac{F}{A}(\rho,T) &=&
 a \rho + b \rho^c + d
\nonumber\\&&
 +\, \tilde{a} t^2 \rho
 + \tilde{b} t^2 \ln(\rho)
 + ( \tilde{c} t^2 + \tilde{d} t^{\tilde{e}} )/\rho \:,
\label{e:fitf}
\eea
where $t=T/(100\mev)$ and $F/A$ and $\rho$ are given in
MeV and $\!\fm3$, respectively.
The parameters of the fits are listed in Table~\ref{t:fit}
($a,b,c,d$ are identical for BHF and FCA by construction)
and are valid in the ranges of density
[$0.05\fm3 \lesssim \rho \lesssim 1\fm3$]
and temperature [$5\mev \lesssim T \lesssim 70\mev$].
The rms deviations of fits and data are better than $1\mev$ for FCA
and $4\mev$ for the BHF procedure.

\subsection{Hot stellar matter}
\label{s:eosT}

For asymmetric nuclear matter,
it turns out that the dependence on proton fraction
can be very well approximated by a parabolic law at zero
\cite{Bombaci94,Burgio10}
as well as at finite temperature \cite{Zuo04,Li21},
\be
 \frac{F}{A}(\rho,T, x_p) \approx
 \frac{F}{A}(\rho,T, 0.5) + (1-2x_p)^2 \frac{F_\text{sym}}{A}(\rho,T) \:
\label{e:parab}
\ee
with the free symmetry energy
\be
 \frac{F_\text{sym}}{A}(\rho,T) =
 \frac{F}{A}(\rho,T,x_p=0) - \frac{F}{A}(\rho,T,x_p=0.5)
\:.
\label{e:fsym}
\ee
Therefore, for the treatment of beta-stable matter,
it is only necessary to provide parametrizations for SNM and PNM,
as by Eq.~(\ref{e:fitf}).
The free symmetry energy determines the composition of stellar matter,
which is governed by both the strong and the weak interaction involving leptons.
We have shown in \cite{Li21} that going beyond the parabolic approximation
affects the results for NS structure only in a very marginal way.

In this article we study exclusively beta-stable
and neutrino-free nuclear matter,
which might not necessarily be a good approximation for merger simulations
\cite{Alford18,Figura21,Hammond21}
and is surely not adequate for supernovae
and the first seconds of life of a proto-NS,
where the effects of trapped neutrinos are essential
\cite{Prakash97,Pons99,Burgio10,Lattimer16}.
In beta-stable nuclear matter containing nucleons and leptons
as relevant degrees of freedom,
the matter composition is determined by imposing chemical equilibrium
and charge-neutrality conditions
at given baryon density $\rho$,
along with baryon charge conservation, i.e.,
\bea
 \mu_n - \mu_p &=& \mu_e = \mu_\mu \:,
\\
 \sum_i Q_i \rho_i &=& 0 \:,
\label{e:beta}
\eea
where the various chemical potentials are obtained from the total
free energy density $f$ according to Eq.~(\ref{e:mui}).
Once the composition $\{\rho_i\}(\rho)$ is found,
the total pressure $p$ and the internal energy density $\eps$ can be calculated
through the thermodynamical relations Eqs.~(\ref{e:eosp},\ref{e:eoss}).

The $p(\eps)$ relation is the essential input for solving the
well-known hydrostatic equilibrium equations
of Tolman, Oppenheimer, and Volkov (TOV) \cite{Shapiro08},
which provide mass $M$ and radius $R$
for a chosen central value of the density $\rho_c$.
The solution of the TOV equations depends on the temperature profile $T(r)$,
as obtained from realistic simulations of astrophysical scenarios
at finite temperature such as
core-collapse supernovae, proto-NSs, and binary NS mergers.
In this study, we continue our simplified analysis of the global effects
of finite temperature on the stability of a NS merger remnant
(see, e.g., \cite{Burgio11,Figura20,Figura21}),
by employing frequently-used isentropic temperature profiles at constant
$S/A=s/\rho=0,1,2,...$ throughout the stellar matter.

The BHF approach provides only the EOS for the bulk matter region
$\rho\gtrsim 0.1\fm3$
without cluster formation,
which is therefore joined with the low-density
Shen20 \cite{Shen20} EOS,
as described in detail in \cite{Burgio10}.
The maximum-mass domain that we are interested in,
is in any case hardly affected by the structure
of this low-density transition region \cite{Burgio10}.
The choice of the crust model can influence the predictions of
radius and related deformability to a small extent,
of the order of $1\%$ for $R_{1.4}$ \cite{Burgio10,Baldo14b,Fortin16},
which is negligible for our purpose.
Even neglecting the crust completely,
NS radius and deformability do not change dramatically \cite{Tsang19}.

\section{Results}
\label{s:res}

\begin{figure}[t]
\vspace{-18mm}
\centerline{\includegraphics[scale=0.34]{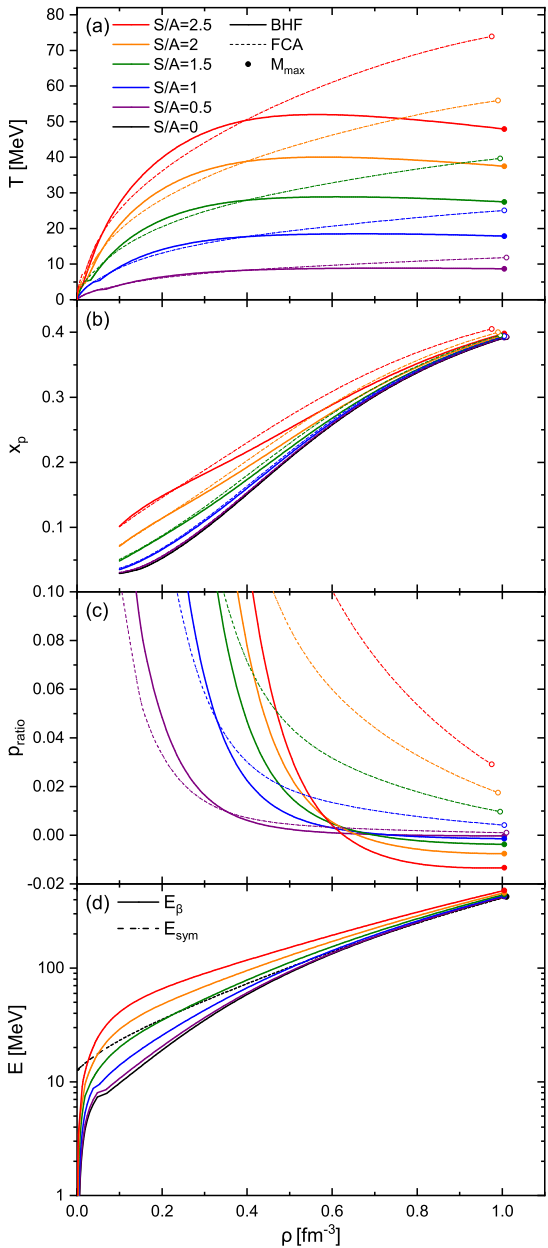}} 
\vspace{-24mm}
\caption{
NS matter properties for fixed specific entropy $S/A$ profiles
as a function of the baryon density:
(a) temperature,
(b) proton fraction,
(c) thermal pressure ratio Eq.~(\ref{e:pr}),
(d) energy per nucleon $E_\beta$
of beta-stable matter and symmetry energy $\esym$ at $T=0$.
All curves end at the central density of their respective $\mmax$ configuration,
indicated by a marker.
Results for BHF (solid curves) and FCA (dashed curves) procedures are compared.
}
\label{f:t}
\end{figure}

In the following we present the results of our numerical calculations
regarding the composition of hot NS matter and the structure of NSs.
One goal is to explore eventual differences between the results
obtained with the FCA and BHF procedures.

\subsection{Hot stellar matter}

Fig.~\ref{f:t} shows several properties of beta-stable charge-neutral matter
as a function of baryon density for several values of fixed entropy $S/A$,
obtained using either the FCA (dashed curves) \cite{Lu19}
or the full BHF procedure (solid curves).
All curves end at their the maximum mass configuration.

Panel (a) shows the temperature profiles.
One notes that the main difference between the FCA and BHF procedures is the
non-monotonic behavior in the latter case.
This is again due to the softening of the EOS at high density
caused by weakened TBF,
see Fig.~\ref{f:fa},
which due to the relation Eq.~(\ref{e:eoss})
yields a much larger entropy for given temperature than in the FCA.
The monotonically rising profiles in \cite{Logoteta21} are similar
to the FCA for the same reason.
In detail, for $S/A=2.5$ the temperatures reach about $50\mev$ at high density
in the BHF case,
and up to $70\mev$ for the FCA,
which are typical values for NS mergers \cite{Figura20,Figura21}
and proto-NSs \cite{Burgio11}.
In the crust region the temperatures vanish,
so that NS radii are well defined for the isentropic profiles.

Panel (b) shows the proton fraction,
which is increased by thermal effects in particular in the low-density regime.
In fact, leptons become rather numerous as a result of Fermi distributions
at finite temperature and,
because of the charge-neutrality condition,
this increases the proton fraction and the isospin symmetry of nuclear matter.
On the other hand,
at high density thermal effects are less important
and do not change the composition appreciably when increasing entropy.
We notice that the proton fraction with the V18 EOS is relatively large,
and at $T=0$ it exceeds at relatively low density $\sim 0.37\fm3$
the threshold value $x_\text{DU}\approx0.13$
for the opening of the direct Urca cooling reactions \cite{Yakovlev01,Burgio21}.
Therefore medium-mass NSs can cool down very rapidly,
in agreement with NS mass distributions,
as illustrated in our recent works \cite{Wei19a,Wei20a,Wei20b}.

In \cite{Wei21} we identified the ratio
of the thermal pressure and the cold pressure,
\be
 \prat \equiv \frac{\pt}{p_0}(\rho,T)
 = \frac{p(\rho,x_T,T) - p(\rho,x_0,0)}{p(\rho,x_0,0)} \:,
\label{e:pr}
\ee
where $x_0$ and $x_T$ are the proton fractions of cold and hot beta-stable matter
\cite{Raithel21,Wei21},
as an important indicator for the strength of thermal effects
on stellar stability.
This quantity is shown in panel (c) of Fig.~\ref{f:t}.
Again important qualitative differences between BHF and FCA results are evident,
namely the BHF thermal pressure becomes negative at certain threshold densities,
also due to the dropping temperature profiles in panel (a).
In our previous analysis \cite{Lu19,Figura20,Figura21,Wei21} we already pointed
out the relatively low thermal pressure in the FCA procedure
due to the fact that beta-stable matter becomes more isospin symmetric at high
temperature.
Now that feature is even more enhanced in the full BHF procedure
as a consequence of the weakened repulsion of the TBF.

For completeness we display in panel (d)
the energy per nucleon of beta-stable matter for the BHF case,
where a moderate dependence on the entropy is clearly visible.
Also the symmetry energy at $T=0$ is shown (dashed curve), the value at saturation being about $32\mev$,
together with a slope parameter $L\approx67\mev$.
A more detailed comparison of the cold BHF EOS with experimental constraints
is given in \cite{Wei20,Burgio21,Burgio21b}. We have also checked that the free symmetry energy
depends only very weakly on the entropy.

\begin{figure}[t]
\vspace{-4mm}
\centerline{\hspace{0mm}\includegraphics[scale=0.37]{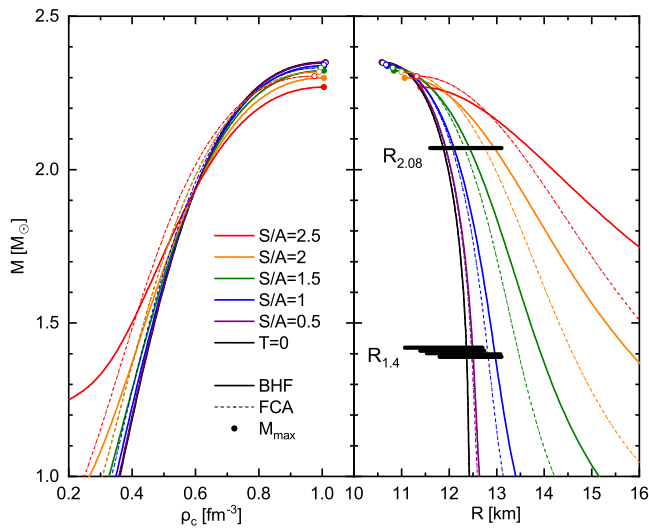}}
\vspace{-8mm}
\caption{
The gravitational NS mass vs.~the central density (left panel)
and the radius (right panel) for several values of $S/A$,
obtained with the BHF/FCA (solid/dashed curves) procedures.
All curves end at their respective $\mmax$ configuration,
indicated by a marker.
The horizontal black bars indicate the limits on
$R_{2.08}$ and $R_{1.4}$
obtained in the combined NICER+GW170817 data analyses
of \cite{Miller21,Pang21,Raaijmakers21}, relevant for the cold EOS.
}
\label{f:mrho}
\end{figure}

\subsection{Stellar structure}

The features of the temperature-dependent EOS
are reflected in Fig.~\ref{f:mrho},
where the gravitational mass vs.~radius and central density relations
are plotted for several values of $S/A$.
Those relations are obtained in the standard way by solving the TOV equations
for beta-stable and charge-neutral matter.
The figure also shows recent mass-radius results of the NICER mission
for the pulsars J0030+0451 \cite{Riley19,Miller19}
and J0740+6620 \cite{Riley21,Miller21,Pang21,Raaijmakers21}.
The combined (strongly model-dependent) analysis
of both pulsars together with GW170817 event observations
\cite{Abbott17,Abbott18}
yields improved limits on $R_{2.08}=12.35\pm0.75\;$km \cite{Miller21},
but in particular on the radius $R_{1.4}$, namely
$12.45\pm0.65\;$km \cite{Miller21}, 
$11.94^{+0.76}_{-0.87}\;$km \cite{Pang21}, and
$12.33^{+0.76}_{-0.81}\;$km or
$12.18^{+0.56}_{-0.79}\;$km \cite{Raaijmakers21},
which are shown as horizontal black bars.
The cold BHF V18 EOS is well compatible with these constraints,
and also its maximum mass $\mmax \approx 2.36\ms$
exceeds the currently known lower limits.
Some theoretical analyses of the GW170817 event
indicate also an upper limit on the maximum mass
of $\sim2.2-2.4\,\ms$ \cite{Shibata17,Margalit17,Rezzolla18,Shibata19},
with which the V18 EOS would be compatible as well.
However, those are very model dependent,
in particular on the still to-be-determined temperature dependence of the EOS
\cite{Khadkikar21,Bauswein21,Figura21}.

Regarding the hot EOS,
in Fig.~\ref{f:mrho} we observe that the maximum masses
obtained with the BHF procedure (full dots)
are slightly smaller than the corresponding FCA ones (open dots),
and both are smaller than the cold configuration (solid black).
But the dependence on entropy is weak.
On the other hand, a more important dependence on the entropy is evident
for the radius of stable mass configurations smaller than two solar masses. One observes a systematic expansion of the hot NS due to thermal effects,
in particular of the less massive objects.
The effect arises mostly from the expansion of the outer core and crust
at finite temperature \cite{Fortin16}.
This explains the somewhat larger radii obtained with the BHF procedure
in comparison to the FCA,
because up to $\rho\lesssim0.4\fm3$ the temperature at fixed $S/A$ is
slighty larger in the former case, see Fig.~\ref{f:t}(a),
which causes a larger expansion.
In fact for very high masses (large densities) the behavior is opposite.
Of course this particular effect is a consequence of assuming idealized
isentropic temperature profiles and might not be relevant in practice.

\begin{figure}[t]
\vspace{-6mm}
\centerline{\hspace{7mm}\includegraphics[scale=0.38,clip]{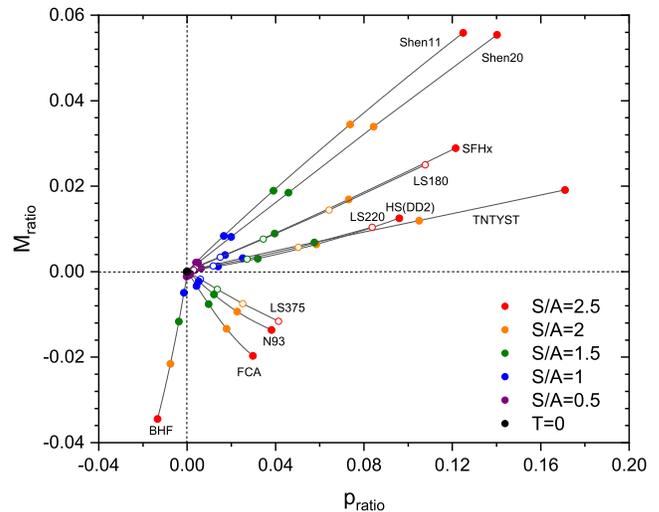}}
\vspace{-6mm}
\caption{
Correlation between
relative change of the maximum gravitational mass,
Eq.~(\ref{e:mr}),
and the pressure ratio at the center of the star,
Eq.~(\ref{e:pr}),
for different $S/A$ and different EOSs.
}
\label{f:dm}
\end{figure}

\subsection{Hot maximum mass and EOS}

In Fig.~\ref{f:dm}
we investigate in more detail the dependence of the maximum gravitational
mass on entropy,
and the relation to properties of the underlying EOS.
For this purpose,
following \cite{Wei21},
we plot the relative change of the gravitational maximum mass,
\be
 \mrat \equiv
 \frac{\mmax^\text{hot} - \mmax^\text{cold}}{\mmax^\text{cold}} \:,
\label{e:mr}
\ee
against the thermal pressure ratio of Eq.~(\ref{e:pr}),
shown in Fig.~\ref{f:t}(c),
taken at the central density of the $\mmax^\text{hot}$ star.
Apart from V18 BHF and FCA results,
we also show in this plot the results obtained for several other
finite-temperature EOSs
that are compatible with the current NS mass observations,
namely yielding $2.1\ms<\mmax<2.4\ms$.
These are the
BHF EOS N93 in FCA \cite{Li08a,Li08b,Burgio10,Lu19,Wei21};
TNTYST \cite{Togashi17} based on the variational APR EOS \cite{Akmal98};
Shen11 \cite{Shen98,Shen11},
Shen20 \cite{Shen20},
DD2 \cite{Hempel10,Typel10}, and
SFHx \cite{Hempel10,Steiner13},
which are based on the covariant density-functional theory.
We also include the set of Lattimer-Swesty EOSs LS180/220/375 \cite{Lattimer91},
which are Skyrme-type EOSs and predict too low or too high $\mmax$,
but share the same finite-$T$ extension to different cold EOSs
and are therefore useful for the theoretical analysis,
as performed in \cite{Wei21}.

\begin{figure}[t]
\vspace{-8mm}
\centerline{\hspace{7mm}\includegraphics[scale=0.38,clip]{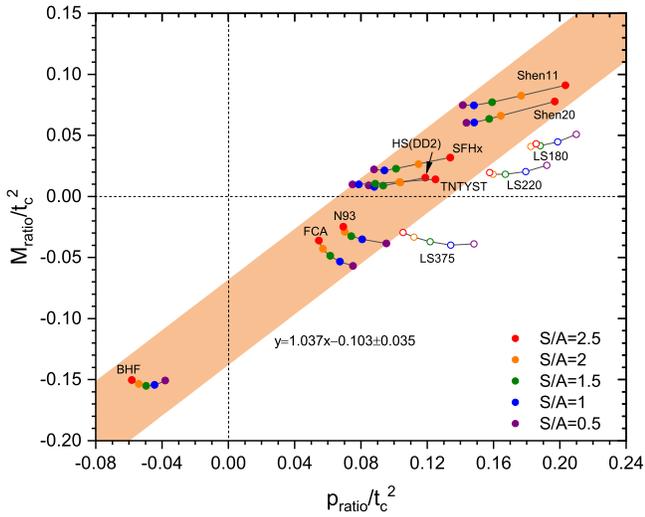}}
\vspace{-6mm}
\caption{
As Fig.~\ref{f:dm},
but $\prat$ and $\mrat$ scaled by the squared core temperature
of the proper stellar profile,
$t_c\equiv T_c/100\mev$.
}
\label{f:dmt}
\end{figure}

For all EOSs one observes a nearly linear relation
between $\mrat$ and $\prat$ with varying $S/A$.
Nearly all sectors of the $\mrat$--$\prat$ plane
are occupied by different EOSs,
apart from $\mrat>0$, $\prat<0$.
Even for the hottest $S/A=2.5$ configuration studied,
the relative change of the maximum mass is limited to less than about 5 percent,
decrease for the BHF EOSs and LS375, and increase for the others.
In particular we notice that the three versions of the LS EOS
produce both positive and negative $\mrat$,
depending on the stiffness of the cold EOS.
In fact, as already found in \cite{Wei21},
the magnitude of the thermal effects depends on the nucleon effective mass
and also on the stiffness of the cold part.
For comparison,
in \cite{Logoteta21} the maximum mass is nearly constant,
in the hottest case $S/A=3$ the increase is not larger than 1\%.

In \cite{Wei21} we found for a fixed $S/A=2$ a linear correlation
between $\mrat$ and $\prat$,
even involving a large number of EOSs that do not respect the
above limits on $\mmax$.
In order to generalize this correlation to arbitrary values of $S/A$,
we note that the dependence of both $\mrat$ and $\prat$
on $S/A$ is approximately parabolic.
This suggests to consider appropriately rescaled quantities
for a `universal relation'.
We found that rescaling both quantities by the square of the central
core temperature $T_c$ of the given fixed-$S/A$ profile
produces a satisfactory correlation
that is displayed in Fig.~\ref{f:dmt}.
For any EOS (respecting the $\mmax$ constraint)
and any thermal condition,
the values of $\mrat/T_c^2$ and $\prat/T_c^2$ are linearly correlated.
In other words, an EOS featuring a certain value of $\prat/T_c^2$
is expected to produce $\mrat/T_c^2$ in a narrow interval,
and vice versa.
For example, a vanishing effect of temperature on $\mmax$, $\mrat=0$,
is expected for an EOS with $0.06<\prat/T_c^2<0.13$,
and a decrease of $\mmax$ for smaller values.
Of course, the proposed correlation should be verified for other
finite-$T$ EOSs once they become available.

\begin{figure}[t]
\vspace{-0mm}
\centerline{\hspace{0mm}\includegraphics[scale=0.39,clip]{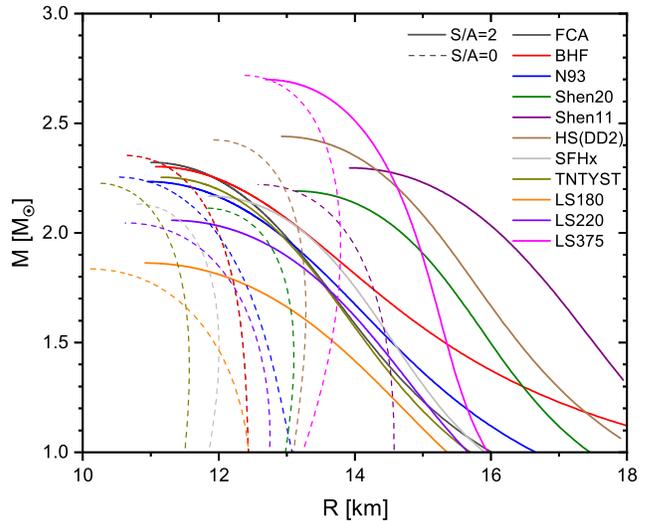}}
\vspace{-3mm}
\caption{
The gravitational mass vs.~the radius for the $T=0$ and $S/A=2$ cases,
for all considered EOSs shown in Figs.~\ref{f:dm},\ref{f:dmt}.
}
\label{f:rm}
\end{figure}

For further illustration,
we show in Fig.~\ref{f:rm} the mass-radius relations for all considered EOSs
at $T=0$ and $S/A=2$.
In the latter case, the radii increase according to the temperature profiles
like Fig.~\ref{f:t}(a)
(see \cite{Wei21}),
as discussed before for the BHF EOSs in relation with Fig.~\ref{f:mrho}.
The maximum mass either increases or decreases for hot stars,
as summarized in Fig.~\ref{f:dm}.

\section{Summary}
\label{s:end}

In conclusion, we presented improved microscopic calculations
of the temperature dependence of the high-density nuclear EOS
in the BHF formalism including realistic two-body and three-body forces.
Results are built upon a cold EOS that fulfills all current observational
constraints regarding in particular NS mass and radius.

The approach goes beyond the FCA by computing
the temperature-dependent s.p.~energy
and also the averaged TBF self-consistently.
The latter effect reduces the
effectively repulsive action of the TBF
due to weakened nuclear correlation functions,
thus renders the thermal pressure negative in high-density beta-stable matter,
and leads to a small decrease of the
maximum gravitational NS mass with temperature,
up to about 2\% for a $S/A=2$ temperature profile.
We also devised a universal relation relating relative changes
of maximum gravitational mass and core thermal pressure
for the currently available set of finite-$T$ EOSs.
In all cases temperature effects on stellar structure
are limited to a few percent, which  
will be difficult to disentangle from future observations of NS merger remnants,
for example.

However,
the finite-$T$ extension of the BHF formalism used here
comprises only temperature effects arising from the Fermi distribution functions
in Pauli operator and momentum integrals,
whereas the input (meson-exchange) two- and three-body forces
are assumed temperature independent.
A consistent inclusion of temperature effects in these quantities
has not been attempted so far
and will be a serious theoretical challenge for the future,
as we have shown that they might change the predictions qualitatively,
in particular regarding increase or decrease of the maximum mass.

In this work our analysis was restricted to idealized isentropic temperature
profiles,
which might not be adequate for realistic astrophysical scenarios.
In this regard also the effect of neutrino trapping was completely
disregarded in this schematic investigation,
which focused on the temperature effects of the strong interaction
in simplified isentropic conditions.
The neutrino contributions to thermal energy density and pressure
might also cause a substantial change of the maximum mass
\cite{Kaplan14,Paschalidis12,Lalit19,Marques17}.
But for a consistent analysis much more detailed simulations are required.

A further important issue to consider is the hypothetical appearance of
strange baryonic matter in the NS core.
A few finite-temperature EOSs including hyperons
are available \cite{Ishizuka08,Burgio11b,Oertel16},
but not all comply with the current $\approx 2\ms$ lower limit on the
maximum gravitational mass of cold non-rotating NSs.
In any case, from many past studies it is known that the presence
of hyperons increases the maximum mass of hot (proto-)NSs,
allowing a possible `delayed collapse' phenomenon during the proto-NS cooldown,
see, e.g., \cite{Prakash97,Nicotra06,Burgio11b,Raduta20} and references therein.

\section*{Acknowledgments}

This work is sponsored by
the National Natural Science Foundation of China under Grant
Nos.~11975077,11475045
and the China Scholarship Council, File No.~201806100066.
We further acknowledge partial support from ``PHAROS'', COST Action CA16214,
and from the agreement ASI-INAF n.2017-14-H.O.

\newcommand{\physrep}{Phys.~Rep.}
\newcommand{\nphysa}{Nucl.~Phys.~A}
\newcommand{\npa}{Nucl.~Phys.~A}
\newcommand{\aap}{A\&A}
\newcommand{\mnras}{MNRAS}
\newcommand{\epja}{EPJA}
\newcommand{\araa}{Annu. Rev. Astron. Astrophys.}
\newcommand{\apjs}{ApJS}
\newcommand{\apjl}{Astrophys.~J.~Lett.}
\newcommand{\rpp}{Rep. Prog. Phys.}
\newcommand{\ppnp}{Prog. Part. Nucl. Phys.}
\newcommand{\plb}{Phys. Lett. B}
\bibliographystyle{apsrev4-1}
\bibliography{fmict}

\end{document}